\documentclass[11pt]{article}
\usepackage[utf8]{inputenc}
\usepackage{url}
\usepackage{hyperref}
\usepackage[backend=biber, sorting=none]{biblatex}
\usepackage[letterpaper,margin=1.7cm]{geometry}

\usepackage{enumitem}

\title{Snowmass'21 Whitepaper - IsoDAR Overview}
\author{J.R.~Alonso, J.M.~Conrad, Y.D.~Kim, S.H.~Seo, M.H.~Shaevitz, J.~Spitz, D.~Winklehner}
\date{March 2022}

\begin{document}

\maketitle

\section{Introduction}

IsoDAR@Yemilab is a unique facility for underground neutrino physics. The system comprises an accelerator-driven $\bar{\nu}_e$ source located next to the Yemilab LSC 2.3 kt detector. Because this facility is first-of-its-kind, it opens new approaches to Beyond Standard Model (BSM) physics searches. The program is most well-known for its capability to perform searches for new oscillation signatures at high statistics in a model-agnostic manner. IsoDAR@Yemilab can definitively resolve the question of $\bar{\nu}_e$ disappearance at short baselines. Beyond this, IsoDAR offers a broad range of searches for new neutrino properties and new particles. The facility uses a state-of-the art cyclotron, that is now fully designed and is undergoing protoyping. Preliminary approval to run at Yemilab in South Korea has led to the completed excavation of caverns. While the accelerator is designed to run underground, IsoDAR accelerators can also be constructed on the surface, allowing this project to contribute to the opportunity for production of life-saving medical isotopes.

The capabilites, technical elements, and deployment studies are well-documented in articles on arXiv, and appear in multiple Snowmass21 whitepapers. Rather than repeat this text,  this whitepaper provides a ``table of contents'' to these documents.

\section{Relevant references on sterile neutrinos}
IsoDAR@Yemilab has unprecendented sensitivity to new physics through oscillations with $\bar \nu_e$. This is due to the single-isotope high-rate $\bar\nu_e$ flux, the well-understood inverse beta decay cross section for detection, and the size of the LSC detector at Yemilab. The oscillation wave-pattern can be studied without reference to any underlying phenomenology from $L/E\sim 1$ to 10 m/MeV. For references, see:
\begin{itemize}[topsep=2pt,itemsep=2pt,parsep=2pt]
    \item Ref.~\cite{NF02} -- NF02 Snowmass'21 Whitepaper on light sterile neutrino searches and related phenomenology, with plots of example oscillation wave-patterns for various popular models for the short baseline anomalies.
    \item Ref.~\cite{PRD} -- A comprehensive article on physics from IsoDAR@Yemilab that includes sensitivities within 3+1 models, plots describing potential to study wavepacket decoherence effects, and examples of oscillation wave-patterns.
\end{itemize}

\section{Relevant references on IsoDAR BSM searches}
IsoDAR@Yemilab is highly sensitive to a number of beyond standard-model scenarios involving both neutrino production, propagation, and interaction, and axion production. The experiment will be able to search for new bosons created at the target, which then decay to neutrino-antineutrinos and subsequently interact in the detector; neutrino wavepacket effects, in which the relevant size of the neutrino wavepacket at creation affects short-baseline oscillation probability; short-baseline oscillations, involving multiple steriles and/or sterile neutrino decay; and non-standard neutrino interactions via $\overline{\nu}_e$-electron elastic scattering. The IsoDAR@Yemilab experimental capability to explore these topics is world-leading for much of the relevant parameter space. For references, see:
\begin{itemize}[topsep=2pt,itemsep=2pt,parsep=2pt]
    \item Ref.~\cite{PRD} -- See above. This article also discusses BSM searches with IsoDAR.
    \item Ref.~\cite{NF03} -- NF03 Snowmass'21 Whitepaper about Dark Sector Studies with Neutrino Beams. It contains a section on Axion Light Particle (ALP) searches with IsoDAR.
    \item Ref.~\cite{NF03_2} -- NF03 Snowmass'21 Whitepaper Beyond the Standard Model Effects on Neutrino Flavor
    \item Ref.~\cite{loyd_alp_upcoming} -- This upcoming publication will contain a detailed description of the IsoDAR sensitivity to axion light particles (ALPs).
    \item Ref.~\cite{conrad_precision_2014} -- Paper on using IsoDAR to collect $\overline{\nu}_e$-electron elastic scattering events and search for non-standard interactions.
    \item Ref.~\cite{bungau:isodar} -- The original IsoDAR sterile neutrino search, published in PRL.
\end{itemize}
    
\section{Relevant references on the IsoDAR Accelerator}
In order to reach the required cw proton beam of 10~mA at 60 MeV, we developed
a compact cyclotron that comprises several innovations. We accelerate H$_2^+$ 
instead of protons, we directly, axially inject via a radiofrequency quadrupole 
(RFQ), and we utilize a collective beam dynamics effect (vortex motion). These 
innovations that are about to be experimentally demonstrated in a 1~MeV/amu test
cyclotron, enable the physics program and the broader impacts of this project. They are
described in detail in teh following references:
\begin{itemize}[topsep=2pt,itemsep=2pt,parsep=2pt]
    \item Ref.~\cite{AF02} -- AF02 Snowmass'21 Whitepaper summarizing the findings of the 
          Snowmass'21 Workshop on High-Power Cyclotrons and FFAs that was held at the 
          request of the Accelerator Frontier (AF) and Acclerators for Neutrinos (AF02)
          conveners.
    \item Ref.~\cite{winklehner2021order} -- A recent publication explaining 
          vortex motion and showing highly accurate particle-in-cell simulations
          demonstrating this effect in the IsoDAR cyclotron and showing clean extraction
          of a 5~mA H$_2^+$ beam (equivalent to 10~mA of protons).
    \item Ref.~\cite{winklehner:rfq, winklehner:nima} -- Publications explaining the
          RFQ direct injection scheme and RFQ design.
\end{itemize}

\section{Relevant references on the underground deployment of IsoDAR}

IsoDAR is collaborating with the Center for Underground Physics Division of the Korean Institute for Basic Sciences for deployment of the IsoDAR neutrino source in close proximity to the planned 2.3 kiloton Liquid Scintillation Counter (LSC) that will be installed in the newly-built Yemilab complex.  The interfacing and deployment of IsoDAR at Yemilab is covered in the following references:

\begin{itemize}[topsep=2pt,itemsep=2pt,parsep=2pt]
    \item  Ref.~\cite{isodar_yemilab} -- Conceptual Design Report with extensive description and illustrations of hardware components, site preparation, cavern, utility, and environmental requirements.
    \item Ref.~\cite{Alonso:2022mup} -- Condensation of CDR prepared for publication in JINST.
\end{itemize}

\section{Relevant references on applications of accelerator to isotope production}

The IsoDAR cyclotron has important, direct applications for medical isotope production. At present, isotopes produced by accelerators in the 50 to 100 MeV energy range come from national laboratories.    The small footprint of the IsoDAR cyclotron makes relatively local production sites feasible and the high rates allow for production well beyond present capabities.   For references see:
\begin{itemize}[topsep=2pt,itemsep=2pt,parsep=2pt]
\item Ref.~\cite{NF07} -- NF07 Snowmass'21 Whitepaper describing the application of IsoDAR technology to isotope production, with a figure showing extraction and beam splitting between targets.
\item Ref.~\cite{alonso:isotopes} -- {\it Nature Physics} review on application of IsoDAR to isotope production
\item Ref.~\cite{waites:isotopes} -- Article in journal for medical researchers on IsoDAR capability.

\end{itemize}

\clearpage
\printbibliography

\end{document}